%% file: template.tex
\documentclass{vgtc}                        

\ifpdf
  \pdfoutput=1\relax                   
  \usepackage{graphicx}                
  \DeclareGraphicsExtensions{.pdf,.png,.jpg,.jpeg} 
\else
  \usepackage{graphicx}                
  \DeclareGraphicsExtensions{.eps}     
\fi%

\graphicspath{{figures/}{pictures/}{images/}{./}} 

\usepackage{microtype}                 
\PassOptionsToPackage{warn}{textcomp}  
\usepackage{textcomp}                  
\usepackage{mathptmx}                  
\usepackage{times}                     
\usepackage[font={small}]{caption}
\usepackage{cite}                      
\usepackage{tabu}                      
\usepackage{booktabs}                  

\usepackage[usenames,dvipsnames]{xcolor}
\usepackage{amsmath}
\usepackage{amssymb}
\usepackage{array}
\usepackage{booktabs}
\usepackage{cancel}
\usepackage{color}
\usepackage{enumitem}
\usepackage{gensymb}
\usepackage{listings}
\usepackage{mathdots}
\usepackage{multirow}
\usepackage{tabularx}
\usepackage{tikz}
\usepackage{siunitx}
\usepackage{soul}
\usepackage{svg}
\usepackage{xspace}
\usepackage{yhmath}

\usepackage{hyperref}
\hypersetup{nolinks=true}

\usetikzlibrary{fadings}
\usetikzlibrary{positioning}
\usetikzlibrary{calc}
\usetikzlibrary{arrows}
\usetikzlibrary{decorations.markings}
\newcommand*{\StrikeThruDistance}{0.15cm}

\tikzset{strike thru arrow/.style={
  decoration={markings, mark=at position 0.5 with {
    \draw [blue, thick,-] 
      ++(-\StrikeThruDistance,-\StrikeThruDistance) 
      --( \StrikeThruDistance, \StrikeThruDistance);
  }},
  postaction={decorate},
}}

\def\LANG{DIVA\xspace}
\def\INSITU{\textit{in situ}\xspace}
\newcommand{\DivaConcept}[1]{\textit{#1}}
\newcommand{\add}[1]{#1} 
\definecolor{codeGreen}{rgb}{0,0.6,0}
\definecolor{codeBlue}{rgb}{0,0,1}
\definecolor{codeRed}{rgb}{0.65,0.11,0.36}
\definecolor{codeGray}{rgb}{0.5,0.5,0.5}
\definecolor{codeMauve}{rgb}{0.58,0,0.82}
\definecolor{codeCyan}{rgb}{0,0.52,0.70}
\newcommand{\ListCodeText  }[1]{{\ttfamily{#1}}}

\def\lstfontsize{\fontsize{6.5}{6.5}\selectfont}
\onlineid{1016}

\vgtccategory{Research}

\vgtcinsertpkg

\title{\LANG: A Declarative and Reactive Language for \INSITU Visualization}

\author{
Qi Wu\thanks{e-mail: qadwu@ucdavis.edu}\\
\parbox{1.8in}{
\scriptsize\centering
University of California, Davis, \\
United States}
\vspace{0.5em}
\and
Tyson Neuroth\thanks{e-mail: taneuroth@ucdavis.edu}\\
\parbox{1.8in}{
\scriptsize\centering
University of California, Davis, \\
United States}
\vspace{0.5em}
\and
Oleg Igouchkine\thanks{e-mail: oigouchkine@ucdavis.edu}\\
\parbox{1.8in}{
\scriptsize\centering
University of California, Davis, \\
United States}
\vspace{0.5em}
\and
Konduri Aditya\thanks{e-mail: konduri.adi@gmail.com}\\
\parbox{1.8in}{
\scriptsize\centering
Indian Institute of Science,
India}
\vspace{-0.5em}
\and 
Jacqueline H. Chen\thanks{e-mail: jhchen@sandia.gov} \\
\parbox{1.8in}{
\scriptsize\centering
Sandia National Laboratories, \\
United States}
\vspace{-0.5em}
\and 
Kwan-Liu Ma\thanks{e-mail: ma@cs.ucdavis.edu} \\
\parbox{1.8in}{
\scriptsize\centering
University of California, Davis, \\
United States}
\vspace{-0.5em}}

\input{contents/00-abstract.tex}

\begin{document}


\firstsection{Introduction}

\maketitle

\input{contents/01-intro.tex}
\input{contents/02-background.tex}
\input{contents/03-design.tex}
\input{contents/04-impl.tex}
\input{contents/05-examples.tex}
\input{contents/08-discussion.tex}

\acknowledgments{The authors wish to thank Martin Rieth at Sandia National Laboratories 
for providing advice, support, and data for this research. 
This research is sponsored in part by the U.S. Department 
of Energy through grant DE-SC0019486.
The work at Sandia National Laboratories was supported by the US Department of Energy, Advanced Scientific Computing Research Office. 
Sandia National Laboratories is a multimission laboratory managed and operated by National Technology and Engineering Solutions of 
Sandia, LLC., a wholly owned subsidiary of Honeywell International, Inc., for the U.S. Department of Energy’s National Nuclear Security 
Administration under contract DE-NA0003525. The work at Indian Institute of Science was supported by the institute's Start-up Research Grant provided to Konduri Aditya.
}

\bibliographystyle{abbrv-doi}

\bibliography{template}



\end{document}

%% file: contents/00-abstract.tex
\abstract{
The use of adaptive workflow management for \INSITU visualization and analysis has 
been a growing trend in large-scale scientific simulations.
However, coordinating adaptive workflows with traditional procedural programming 
languages can be difficult
because system flow is determined by unpredictable scientific phenomena, which often
appear in an unknown order and can evade event handling.
This makes the implementation of adaptive workflows tedious and error-prone.
Recently, reactive and declarative programming paradigms have been recognized as 
well-suited solutions to similar problems in other domains.
However, there is a dearth of research on adapting these approaches to \INSITU visualization and analysis.
With this paper, we present a language design and runtime system 
for developing adaptive systems through a declarative and reactive programming 
paradigm. We illustrate how an adaptive workflow programming system is implemented using our 
approach and demonstrate it with a use case from a combustion simulation.
} 

%% file: contents/01-intro.tex

Scientific simulations running on petascale high-performance computing
(HPC) platforms such as Summit~\cite{SummitSuperComputer}
can easily produce datasets at scales beyond what can be efficiently
processed.
Therefore, scientists often have to compromise the
allocation of strained resources, such as I/O bandwidth, 
to maximize the overall efficiency of simulations
~\cite{InSituSummary, 5307638}.
An ideal solution is to manage the system workflow adaptively
through trigger-action mechanisms
(which are termed ``\INSITU triggers'' in some literature~\cite{InSituTrigger, bennett2016trigger})
because they can prioritize the allocation of these strained resources to the 
most interesting phenomena as they emerge
~\cite{AdaptiveWorkFlow}.
However, 
adaptive workflows are usually more difficult and error-prone to program
because they are \textit{reactive} applications rather than 
\textit{transformational}~\cite{Frp-Continued}.
Control flows in adaptive workflows are often driven by evolving simulation 
outcomes and event sequences that cannot be predicted in advance.
Therefore, domain scientists
must anticipate 
all possible scenarios in advance
and create sophisticated rules to dynamically
trigger actions in response to the events~\cite{bainomugisha2013survey}.
Moreover, because the quality of adaptive workflows often relies heavily
on the accuracy of trigger conditions,
having an \INSITU infrastructure that simplifies
the customization process is important~\cite{InSituSummary, InSituTrigger}.

A common approach to implementing reactive systems is through synchronous 
dataflow programming models~\cite{SIGNAL, LUSTRE, LucidSynchrone}. 
In such models, programs are formed by wiring primitive processing elements and 
composition operators into a directed graph structure. The dataflow programming 
model provides a natural form of modularity for many applications.
VTK~\cite{VTK} also partially adopted this approach, 
however, VTK is designed for programming unidirectional visualization pipelines,
and provides limited support for highly dynamic dataflows.
Moreover, the synchronous dataflow model is somewhat difficult to use and does not 
always lead to modular programs for large scale applications when control 
flows become complicated~\cite{YampaArcade}.
Functional reactive programming (FRP)~\cite{YampaArcade, Fran, Frp-Continued, 
Frp-Refactored} further improved this model by directly treating time-varying 
values as first-class primitives. This allowed programmers to write reactive programs
using dataflows declaratively (as opposed to callback functions), 
hiding the mechanism that controls those flows under an 
abstraction.
By enabling such a uniform and pervasive manner to handle complicated data flows,
applications gain clarity and reliability.

Through our work, 
we demonstrate how FRP abstractions can be used to better assist
adaptive \INSITU visualization and analysis workflow creation,
using a domain-specific language (DSL) we created called \LANG.
This language consists of two components:
an FRP-based visualization specific language and 
a low-level C++-based dataflow API. 
Rather than replacing existing \INSITU infrastructures,
it aims to work as a middle layer which can extend
existing systems such as VTK~\cite{VTK}. 
Through the description of our design, we emphasize the key principles that ensure a correct implementation of the FPR abstractions in a parallel environment. 
These principles can also be applied to \INSITU systems that are aimed at enabling declarative and reactive programming, without adopting \LANG itself.
Our design provides four primary benefits:

\textbf{Simple}.
Traditionally, \INSITU infrastructures employ callback functions 
to handle separated programming stages.
Useful callback functions include initialization, 
per-timestep execution, finalization, and feedback loops~\cite{loring2018python}.
As such, a single data dependency could be broken into multiple pieces that are
controlled by interleaving control flows, making it hard to read and maintain.
If an operation requires information from multiple timesteps, the handling of static 
storage will also be involved, making the implementation even more complex. 
Declarative programming simplifies 
these tedious and redundant
low-level tasks by placing more of the burden on the tool developers.
This allows the user to focus on specifying the results they desire.
Meanwhile,
reactive programming offers the ability to automatically
coordinate data dependencies and propagate changes from inputs to outputs.
This model is commonly used to greatly simplify the handling of time-varying signals,
such as events triggered by human interaction. Since adaptive workflows
similarly specify the reaction of a system to time-varying signals, we
believe reactive programming is also a good solution for coordinating them.

\textbf{Extensible}.
By providing a fully programmable interface and a low-level C++ API, 
\LANG 
allows flexible extension through the development of new custom modules. 

\textbf{Portable}. 
The low-level API also makes the integration with existing visualization 
infrastructures easier, 
by only asking for a short binding implementation; this can typically be just 
one source file. Thus, \LANG can be used as a wrapper to
enable declarative and reactive programming for these infrastructures.
This also means that workflows written in \LANG for one infrastructure can be 
easily reused in a different infrastructure if the proper 
bindings are supplied for 
both of them.

\textbf{Dynamic}.  
\LANG's API resolves linking through dynamic loading; therefore,
adding or updating linked libraries does not always require recompilation nor 
restarting.
This feature can be useful for scientific simulations on HPC systems, because 
allocations for these simulations are usually limited; removing unnecessary restarts 
allows for more useful simulation work to be done and gives a higher 
chance for scientific discoveries.

In this paper, we
first begin with a summary of related works. 
We then describe
the principles and features 
of our language 
with  a set of examples.
We also provide a set of examples to demonstrate how our design meets 
the needs of \INSITU visualization and analysis,
and how it supports our assertion 
that declarative and reactive models are good approaches 
for adaptive workflow management in this domain.
Next, we discuss implementation details of our language.
Finally, we showcase the use of \LANG to program an adaptive \INSITU workflow for a leading edge simulation code running on the Summit supercomputer.

%% file: contents/02-background.tex
\section{Background and Motivation}



We begin with an overview of recent advances in 
reconfigurable 
\INSITU workflows, and a history of functional reactive programming.
We then compare existing methods in our domain with the FRP model.
Finally, we conclude with a discussion about how our approach 
is suitable for better assisting adaptive workflow design.

\subsection{Reconfigurable \INSITU Workflows}


We review recent research in reconfigurable \INSITU workflows from two categories that are particularly relevant to visualization: 
having a ``human-in-the-loop'' with a focus on improved interactivity, and
automatically looking for regions of importance during the simulation.
In the first category, many successful infrastructures for \INSITU visualization
and analysis, such as Libsim~\cite{Libsim} and Catalyst~\cite{Catalyst},
support interactive exploration during the simulation.
However, even with these infrastructures, tasks like searching for infrequent 
scientific phenomena can still be challenging, 
because 
scientists do not always know which aspects of the simulation to focus on.
In the second category, 
many have attempted to automate the search for important regions for in-depth 
analysis, visualization, and storage~\cite{bennett2012combining, 
morozov2013distributed, nouanesengsy2014adr}. Notably, recent works following in this 
direction usually involve defining indicator functions known as ``\INSITU triggers'' 
for characterizing features. 
\add{These triggers can be domain-agnostic algorithms such as data reduction, 
aggregation, statistical analysis, and machine learning~\cite{zhou2018key,
ling2017using, malakar2016optimal, myers2016partitioning, banesh112018change},
or domain-specific routines that require special knowledge from domain-experts
~\cite{AdaptiveWorkFlow, bennett2016trigger, zhao2009simulations, 
ullrich2017tempestextremes}.}
Through this approach, a system could automatically focus resources on
the most interesting phenomena from the simulation, but only if the triggers 
are well-designed.
However, manually designing and implementing these triggers can be tedious 
(e.g., \add{by directly working on an infrastructure's source code}) 
or error-prone (e.g., misuse of algorithms).
\add{Tools to simplify} the development and composition of \INSITU triggers into 
workflows are \add{much needed.}
\add{Recent} work by \add{Larson et al.} demonstrates a flexible \add{interface
for creating \INSITU triggers in the ASCENT~\cite{InSituTrigger} infrastructure.}
\add{Our work is inspired by this.}
However, with \LANG, we contribute a complete DSL for programming trigger-based dynamic workflows. 
\add{In this language, fine-grained \INSITU triggers are automatically generated 
based on user-specified data dependencies and high-level constraints. 
This approach frees users from manually writing every trigger
and allows them to create potentially better and more reliable workflows.}




\subsection{Functional Reactive Programming}

Functional reactive programming (FRP) is a declarative programming paradigm for
working with time-varying values.
Particularly, FRP defines time-varying values as \textit{signals}, which 
conceptually can be viewed as functions that include time as a parameter. 
In \INSITU workflow management, such \textit{signals} can represent the
outputs from a simulation~\cite{Frp-Refactored}.
Although there are many variations of FRP focusing on different applications,
they all fall into two main branches:
classical FRP~\cite{Fran} and arrowized FRP~\cite{Frp-Continued}.

Classical FRP was first proposed for creating interactive animations.
It introduced notations called \textit{behaviors} (another name for signals) and 
\textit{events}, to represent continuous time-varying values and sequences of 
time-stamped values respectively.
Although initially focused on animation, it inspired many later works of broader 
scope due to its elegant semantics~\cite{Elm-Thesis}. 
However, because it is a denotational model which does not restrict the length of a 
signal stream that can be operated on, classical FRP programs can have a high memory
footprint and long computation times~\cite{Elm-Thesis}.

Arrowized FRP (AFRP)~\cite{Frp-Continued} aims to resolve the space and time
leaks without losing the expressiveness of classical FRP. 
Instead of working with signals or similar notations directly, 
AFRP focuses on manipulating causal functions between signals, 
connecting to the outside world only at the top level~\cite{Frp-Refactored}.
However, 
programs written in AFRP can still suffer from issues like global delay or 
unnecessary updates, depending on their implementations~\cite{Elm-Thesis}.

Real-time FRP (RT-FRP)~\cite{RT-FRP}, event-driven FRP (E-FRP)~\cite{E-FRP} and 
asynchronous FRP (e.g., the original version of Elm)~\cite{Elm} 
are other variations developed to optimize classical FRP. 
RT-FRP introduced a two-tiered language design for reactive programming;
It uses a closed, unrestricted language to give direct operational semantics, making
it possible to measure computational expenses~\cite{RT-FRP}, along with a simpler 
and more limited reactive language for manipulating signals~\cite{RT-FRP}.
This separation makes it much easier to prove properties of RT-FRP, 
at the expense of expressiveness~\cite{Elm-Thesis}.
Similarly, E-FRP makes the assumption that signals are all discrete.
This property makes it suitable for intensive event-driven applications,
including interactive visualization~\cite{Vega, Vega-Lite, meyerovich2009flapjax, 
cottam2008stencil, satyanarayan2014declarative}. 
Asynchronous FRP allows programmers to explicitly enable asynchronous event 
processing, and thus enables efficient concurrent execution of FRP programs. 
This ensures that the responsiveness of the user interface will not be affected by 
long-running computations~\cite{Elm}. Implementations of asynchronous FRP can also 
be found in many imperative languages through Reactive Extensions~\cite{reactivex}.


\subsection{Dataflow Model for Visualization}

There is a rich 
literature~\cite{AVS, OpenDX, SCI:SCIRun, VTK, VisTrails, Voreen, Inviwo,Decaf,
Damaris,Henson} on 
using dataflow models to 
realize configurable visualization systems.
In these frameworks, workflows are represented as simple pipelines or directed 
graphs, with nodes representing low-level visualization components 
(also called functions, filters, modules, or processors).
Data is processed hierarchically as it flows through components to form a 
complete workflow~\cite{Min-V3D}.
Even though this method works very well for traditional post hoc scenarios,
there is still room for improvements in terms of 
abstraction and efficiency when adaptive \INSITU workflows are considered.
From the perspective of abstraction, these frameworks do not supply 
first-class representations for time-varying primitives. 
This makes the management of time-varying storage manual and tedious.
From the perspective of efficiency, because \INSITU workflows are expected 
to be executed repeatedly for a large number of timesteps, 
lazy evaluation can help to avoid needless recomputation.
However, many popular visualization frameworks lack 
system-wide support for lazy evaluation, 
leaving it to the developer to avoid unnecessary computation.
This makes the quality of module implementations 
a more important factor for efficiency, and makes it difficult for novice 
programmers to implement efficient modules inside
such frameworks.

\begin{figure*}[tb]
  \begin{center}
  \includegraphics[width=\linewidth]{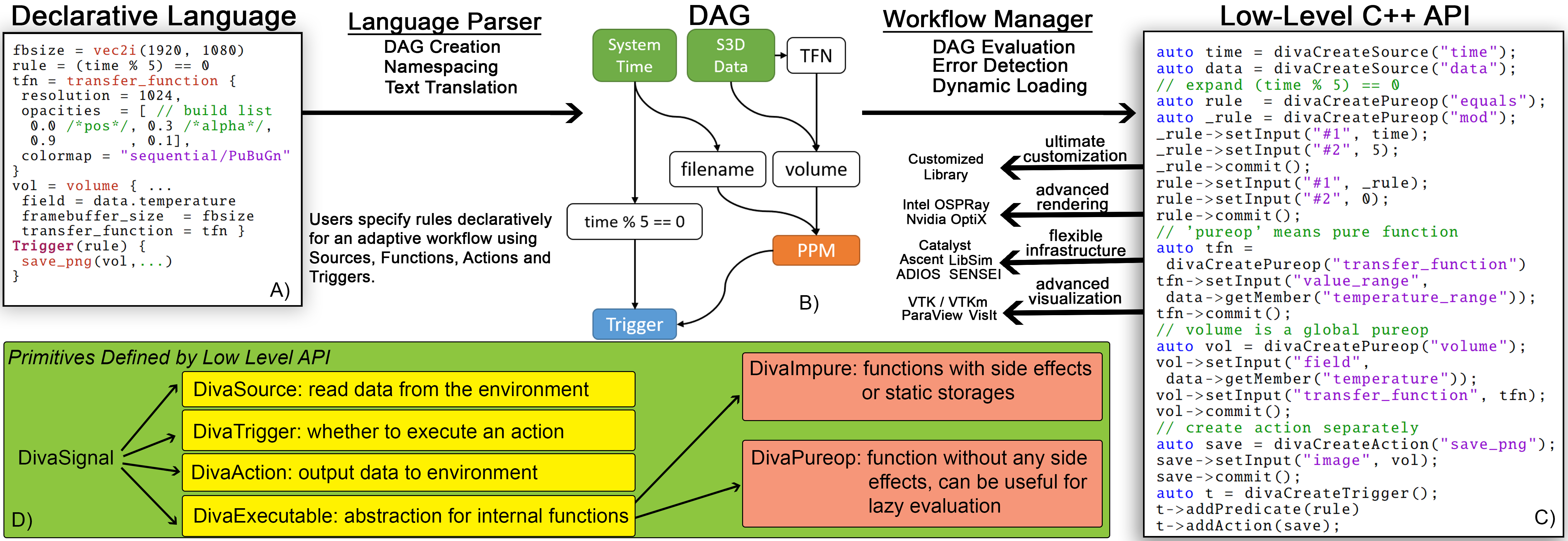}
  \caption{\label{fig:diva-impl}A \LANG program is processed through three layers.
  Users typically specify their program using the declarative interface (left);
  then the language parser will translate it into an internal DAG representation;
  this representation will then be interpreted into a low-level dataflow API for
  execution.
  A) A \LANG program computes a volume rendering for every 5
  timesteps, and saves the rendering on disk. 
  B) The same program in the DAG representation.
  C) The same program in the low-level API.
  Because the C++ API is not declarative, in part C), statements have to be
  executed in order.
  Moreover, because C++ does not track data dependencies automatically, 
  all variables declared in C) should be wrapped by lifting operators 
  (e.g., \ListCodeText{divaCreateSource}).
  D) The hierarchy of primitives defined by the low-level dataflow API: 
  \add{All values in \LANG are signals; values depending on external inputs are sources;
  values returning to the environment are actions (e.g., a saved image file); triggers
  are special primitives that decide which actions to compute based on predicates; 
  rest values are internal to the workflow and are represented by either 
  pure (i.e., \ListCodeText{DivaPureOp}) or impure functions 
  (i.e., \ListCodeText{DivaImpure}).} 
  \vspace{-3.8em}}
  \end{center}
\end{figure*}

\subsection{Programmable Interfaces for Visualization}



By 
directly incorporating low-level APIs (e.g., C++ or Fortran),
users can always create visualizations and analyses using arbitrary algorithms,
and fully utilize modern architectures for cutting-edge techniques such as
ray-tracing~\cite{ospray}, and heterogeneous parallelism. 
However, novice programmers are prevented from
doing so due to the complexity and difficulty in low-level 
programming.
End users often prefer to compose the low-level modules using high-level dynamic languages like Python; and in this field, almost all major \INSITU infrastructures provide this support, by employing one of two 
approaches~\cite{doi:10.1002/cpe.3125, ADIOS, loring2018python}.
In the first approach, the capabilities are exposed through wrapped objects or
functions that can be configured using 
Python~\cite{Libsim, Catalyst, ParaView, VisIt, Ascent, VTKm}. 
In the other, the infrastructure embeds user-supplied high-level codes directly 
into its pipeline, providing users with direct access to the simulation data and the 
ability to analyze natively in the high-level 
language~\cite{SENSEI, loring2018python}.
However, Python has
a major drawback when it comes to supporting FRP features. 
Because Python is designed for a much wider range of tasks than just programming adaptive \INSITU workflows,
it makes fewer assumptions about data abstraction and data dependencies; the implementation of features 
like lazy evaluation and dependency tracking 
often rely on wrapper functions and special inheritance patterns, 
which can be complicated to use and increase the verbosity of the program.

%
%
DSLs use specialized grammars that are tailored for particular domains. 
Several DSLs have been designed for visualization and 
analysis over the years, including Data Explorer~\cite{lucas1992architecture}, 
Scout~\cite{mccormick2007scout}, Diderot~\cite{chiw2012diderot, kindlmann2015diderot}, 
ViSlang~\cite{rautek2014vislang}, Mathematica~\cite{Mathematica}, MATLAB~\cite{matlab} etc.
Such DSLs enable users to express programs in domain-specific semantics that they are
familiar with, and hide more complicated implementation details~\cite{Min-V3D}.
A well-designed DSL can greatly simplify the learning process. 
%
%
A declarative design can further improve a DSL
by introducing only declarative grammars for system configurations.
It also improves usability 
by 
not only providing a large number of domain-specific built-in functions,
but also introducing new language abstractions to
hide underlying execution models~\cite{satyanarayan2014declarative}.
Because FRP itself is a variant of declarative language (for handling time-varying
data), implementing FRP abstractions using a declarative DSL approach is
straightforward.
Several information visualization toolkits have already adopted 
declarative~\cite{Protovis, ggplot2, D3}
and reactive~\cite{Vega, Vega-Lite}
grammars for specifying visual encodings.
Some of these have become very popular due to their ease of use and flexibility.
These works have inspired further work to improve usability in other areas, including high-performance information visualization 
systems~\cite{Kelv-P4},
and for the configuration of complex GPU shader pipelines for advanced volume
rendering of scientific data~\cite{Min-V3D}.
Our work is inspired by these existing designs; however, we focus on using a declarative and reactive programming model to correctly handle parallel
\INSITU visualization and analysis workflows.

\subsection{Motivation for a New Language}
Since it takes time to learn a new programming language, embedding our system within an existing language would have some advantages.
However, we believe that, for programming complicated adaptive \INSITU workflows,
\LANG is favorable, for the following reasons:
First, data abstractions introduced by FRP enable not only
declarative programming, but also systematic tracking of time-varying
data dependencies. 
They also create a new way for users to think of the data (in terms of
time-sequences rather than iteratively updated values). 
Implementing this abstraction in C++ or Python is possible, 
but that would require a new set of data wrappers to be provided.
When programming using this API, they would also need to manually lift 
all the native types provided by the host language using these wrappers.
This process alone might significantly steepen the learning curve.
Second, workflows programmed in \LANG might be more flexible, better to optimize, 
and easier to debug, as the language compiler knows exactly what users write. 
For example, in a well-defined DSL environment, users only need to write down
essential information about the algorithm; technical details such as updating
the DAG and building the dependency diagram can be computed automatically.
Thirdly, code written in an actively developed DSL can be more portable across platforms,
because DSLs are usually simpler and easier to port (due to their limitations and simplicity).
For example, it is fairly easy to rewrite our \LANG system in JavaScript and enable 
web-based visualization. However, porting Python or C++ to browsers can be
challenging compared to rewriting the workflow entirely.
Finally, we do observe that 
it might be difficult to 
fully understand \LANG's methods, 
but we believe that these difficulties are coming from the new abstractions and concepts
being introduced, rather than the syntax of the language itself. 
\LANG's syntax shares many features from popular high-level languages including 
Python and Lua. Thus, understanding \LANG's syntax should not be a challenging task.

%% file: contents/03-design.tex


\section{The \LANG Language Design}\label{sec:design}

We implemented \LANG through three components (as shown in~\autoref{fig:diva-impl}): 
the language parser for translating \LANG workflow specifications into 
DAG representations; the workflow manager, which is in charge of managing system 
resources, detecting logical errors, and making decisions about how to execute the 
DAG;
and the C++ dataflow API for executing low-level components.
In this section, we illustrate our design decisions using examples and
discuss the advantages of these choices through comparison with code written 
in other models.

\subsection{Signal Abstraction}\label{sec:design:signal}


\begin{lstlisting}[float=t!,%
caption={Code comparison between \LANG and Python.
Both programs render streamlines that are computed by integrating the gradient
of an input volume. By introducing the signal abstraction, \LANG can automatically
remember the dependencies between variables. As it is optimized using lazy 
evaluation, it only updates variables that are changing. Therefore, users
do not have to explicitly compute the ``seed'' in the initialization function.
The signal abstraction also automatically considers time-varying objects
as sequences, thus there is no need to manually create static storage, such as 
``pair'' and ``hist'', nor to manipulate them iteratively. Note that, 
syntactically \LANG can use either ``\{\}'' or ``()'' to declare nodes with
parameters.%(Full-size code snapshots can be found in the supplementary material.)
\vspace{-3em}},label={fig:code-signal}]
/* --- DIVA --------------------------------------------- */
pair = ^window^(data.HeatRelease, 2 ~...~)
grad = ^gradient^(pair, data.ldims) // compute gradient
// record gradients for every 10 timesteps
hist = ^window^(/*trigger=*/time%10==0, grad, 100,~...~)
// integrate the streamline
seed       = ^random_3d^(^vec3f^(0), data.ldims)
streamline = ^integrate_line^(seed, hist)
// render and save images when the window is full
img = ^line^(streamline,~...~)
Trigger(time%1000==0) { save_png(img, "img-"+str(time)) }

/* --- Python ------------------------------------------- */
pair, hist, seed = [], [], None
def Init():
  seed = ^random_3d^(^vec3f^(0), data.ldims)
def Process(time):
  pair.append(data.HeatRelease)    
  if (^len^(pair) > 2): pair.pop(0)
  if (time % 10 == 0):
    grad = ^gradient^(pair, data.ldims,~...~)
    hist.append(grad)
    if (^len^(hist) > 100): hist.pop(0)
  if (time % 1000 == 0):
    streamline = ^integrate_line^(seed, hist)
    img = ^line^(streamline,~...~)
    ^save_png^(img, "img-" + ^str^(time))
\end{lstlisting}


\LANG adopts the abstraction of signals from classical functional reactive programming (FRP) to represent
time-varying values. Specifically, a signal of type $\alpha$ in \LANG, denoted as
$\widehat{\alpha}$, can be considered as a function from time to typed values:
\begin{equation*}
  \widehat{\alpha}: \text{Time} \to \alpha
\end{equation*}
In this formulation, time refers to the simulation timesteps, and can be 
represented as a non-negative integer.
For the sake of simplicity, \LANG assumes that all variables are treated as 
signals implicitly and they cannot be deleted or modified upon construction.
\LANG also assumes that new signals can only be created using \DivaConcept{pure functions}, 
where the ``pureness'' is defined conventionally through referential 
transparency~\cite{russell1997principia}. However, there are two exceptions, 
which are discussed in \autoref{sec:design:impure}.
These simplifications make it possible for our system to systematically resolve the 
evaluation schedule for signals and 
correctly update their values when necessary.
It also allows users to process time-sequences directly using array-like 
operations rather than using iterative algorithms.
The effects of the signal abstraction in \LANG can be found in 
\autoref{fig:code-signal}, where a code comparison between \LANG and Python is
demonstrated. In this example, both programs compute and render streamlines
from volume gradients. Because the computation of gradients and streamlines
requires data from multiple timesteps, developers need to explicitly manipulate multiple 
static storages; to avoid unnecessary recomputations, 
they should also seed streamlines
in the initialization function (e.g., \ListCodeText{Init}). 
In \LANG, these steps are not needed. 
Users of \LANG 
can therefore focus less on program execution (e.g., when to generate the initial seedings for streamlines)
and more on the specification of 
desired results (e.g., seedings are the initial condition for the line integral).

\subsection{Language Structure}
\label{sec:design::two-tier-design}




Similar to RT-FRP, \LANG uses a two-tier language design to provide tractable 
notions of computational costs. In particular, there is a restricted reactive 
language for declaratively manipulating signals, and an unrestricted low-level 
imperative API for implementing details of the reactive abstraction.

The reactive language is a combination of classical FRP abstractions and 
the dataflow model used by VTK for constructing visualization pipelines.
However it enriches the use of directed acyclic graphs (DAGs)
and expression of the data flow through \textit{source-filter-mapping},
by replacing variables with signals.
This enrichment makes \LANG's reactive programming interface more suitable for 
writing components that require inputs from multiple timesteps,
because users are freed from the manual handling of global or static storage.
Moreover, in this language, signals and their dependencies are conceptually 
represented as parameterized nodes and directed links.
Based on the location of a node in the DAG, nodes are classified as a
\DivaConcept{source}, \DivaConcept{function}, or \DivaConcept{action}.
\textbf{\DivaConcept{Sources}} are root nodes in the graph for representing 
predefined signals (e.g., simulation outputs) that initiate data flows.
\textbf{\DivaConcept{Functions}} are internal nodes requiring both inputs and outputs
for computing intermediate values within the workflow.
Because all values in \LANG's declarative interface are signals, 
\DivaConcept{functions} are essentially constructors for signals.
As a result, most \DivaConcept{functions} cannot produce side effects.
\textbf{\DivaConcept{Actions}} are terminating nodes in the graph, defining how 
the workflow interacts with the external environment. They are generalizations of 
VTK's mappers, because they can not only map visualizations to display devices, but 
also conduct tasks like data storage and \INSITU steering\footnote{A feedback mechanism 
that allows the modification of the simulation based on visualization/analysis 
outputs while the simulation is running.}. They do not return values
back to the workflow, and their implementations are expected to have side effects.
In addition to the three categories mentioned above, \LANG also introduces a 
special type of node --- \textbf{\DivaConcept{triggers}}, which are 
\add{higher-order}\footnote{A function that takes one or more functions as arguments 
or returns a function as its result.}
functions for signals of the following type:
\begin{equation*}
  \text{Trigger}~A: (\widehat{\text{\ttfamily bool}},~A) \to \text{Maybe}~A
\end{equation*}
where $A$ refers to an \DivaConcept{action} and {\ttfamily Maybe}$~A$ represents a
type that can return either A or nothing. 
They are responsible for dynamically controlling the 
execution of \DivaConcept{actions} based on a Boolean signal: 
when the signal evaluates to true, the corresponding \DivaConcept{action} is
returned and executed; signal evaluates to false,
the execution is skipped.
\autoref{fig:diva-impl} A, shows an example where the ``temperature''
field is rendered once every 5 timesteps.

The low-level interface for \LANG is implemented in C++ 
following a traditional dataflow model.
This API is intended for implementing all the reusable modules that can be called from \LANG,
rather than implementing the workflow itself.
Therefore, users of \LANG do not have to directly use this API.
Because the low-level API is not declarative, \add{code translated into this
API is much more verbose (\autoref{fig:diva-impl}C) 
compared to code written declaratively (\autoref{fig:diva-impl}A).}
However, with this API, \LANG can be extended easily and flexibly,
by taking advantages of
three features.
First, the 
low-level runtime provides a 
simple but generic 
command pattern API. Although simple, this API fully implements 
the reactive language's abstractions. For example, 
generic classes for signals, actions, and \DivaConcept{pure} and \DivaConcept{impure functions}
are provided for programming different features as shown in
\autoref{fig:diva-impl}D;
within those classes, developers can manipulate parameters and return outputs
using functions like \ListCodeText{setInput}, 
\ListCodeText{addAction}, and \ListCodeText{setValue}.
Second, the API provides detailed indications about different stages of computation, 
such as initialize, and commit (which resolves types), and execute.
This allows developers to better optimize their implementations when they are trying
to extend \LANG with customized modules.
However, this does not mean that all of the low-level development burdens are
thrown to developers;
instead, features such as lazy-evaluation 
(see \autoref{sec:impl}) are implemented system wide in base classes, 
making them immediately available to all extensions.
%
Third, through our low-level API, developers can register custom types, that can 
then be used directly in the reactive language layer. 
This helps make it easier to integrate existing, 
well-optimized libraries.
Finally, integrating new packages into \LANG often does not require 
a full code recompilation.
Instead, \LANG uses dynamic loading to look for 
registered function symbols during runtime.
As long as the added packages are 
compiled with the core components and 
placed correctly,
the \LANG back-end system can automatically link them, 
even while the simulation is running.
This allows incremental development 
without restarting a simulation.

\begin{figure}[tb]
\begin{center}
\includegraphics[width=\linewidth]{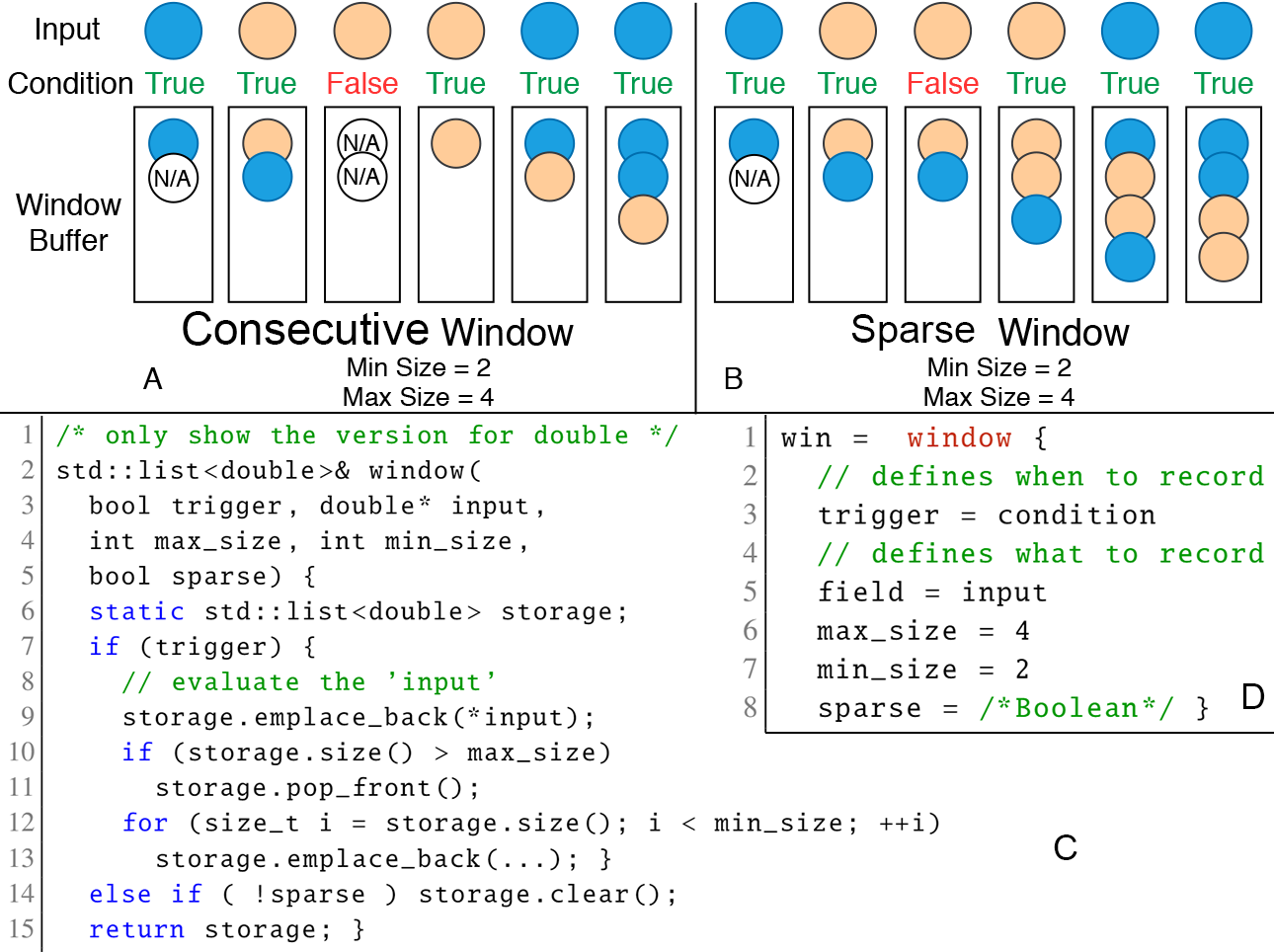}
\end{center}
\vspace{-1em}
\caption{Window is a special
 \DivaConcept{impure function} that collects a signal's historical values
into an array at timesteps where the ``condition'' is true.
A window can be constructed in sparse mode, or consecutive mode. 
In the sparse mode, up to ``max\_size" values are collected, even if they appeared 
sparsely in time. In
the consecutive mode, values that appear within an interval of consecutive time
steps are recorded, and when the condition becomes false, the window is cleared. As 
an example, a consecutive window can track history for a period time while a 
phenomena is active, then automatically reset when the phenomena subsides, and then 
begin tracking the phenomena again if it becomes active another time. 
\vspace{-2.5em}}
\label{fig:window}
\end{figure}
  
\subsection{Impure Functions}\label{sec:design:impure}

As discussed previously, programs written in a classical FRP language have
unrestricted access to a signal's history. Thus a correct classical FRP 
implementation 
will have to to track all signal values automatically. Although this method has been
proven to be fast enough for many applications~\cite{RT-FRP}, holding extra copies
of the simulation data might not be suitable for \INSITU workflows because of
the high expense in terms of processing time and memory.
To solve this issue, \LANG introduces dedicated  \DivaConcept{impure functions} 
for accessing values across time, while other \DivaConcept{functions} are prohibited
from doing so. 
Those \DivaConcept{impure functions} are specialized constructors for creating
stateful signals (i.e., signals with static storage).
During the construction, a finite description of the computational cost 
must be presented explicitly, which effectively prevents users from 
accidentally writing programs that can produce large memory footprints unexpectedly.

\begin{figure}[b]
\begin{center}
\includegraphics[width=\linewidth]{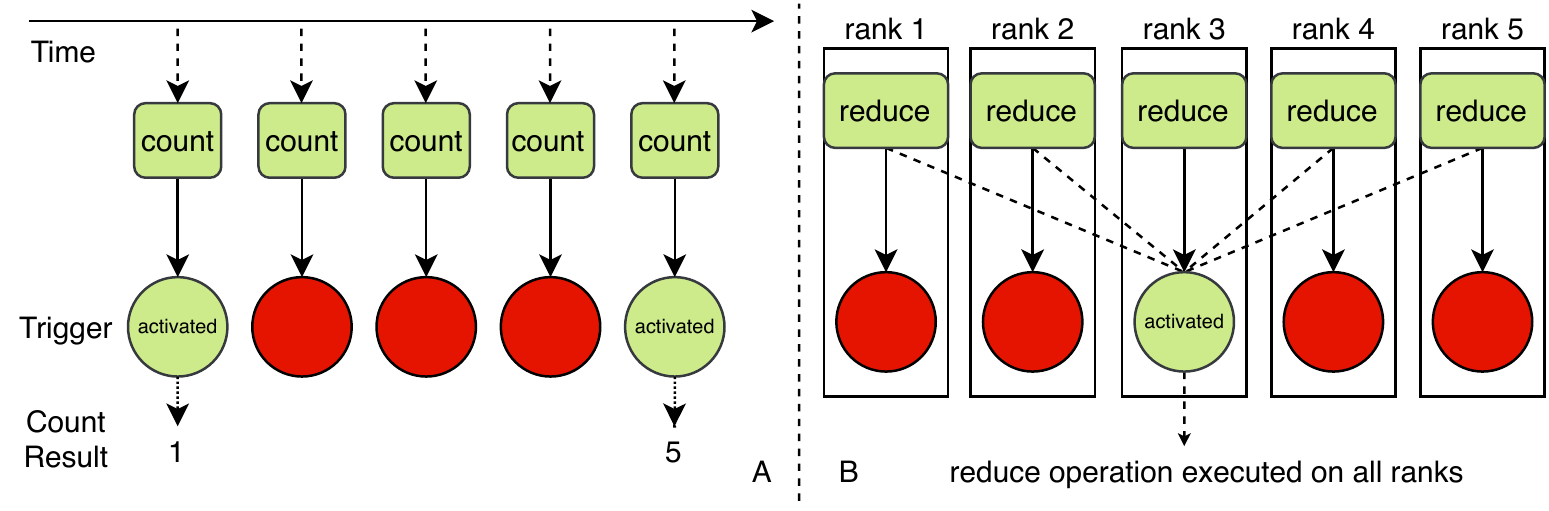}
\end{center}
\vspace{-1em}
\caption{Special rules are used to control the evaluation of 
 \DivaConcept{impure functions} and global synchronizations.
A) As a minimal example, the  \DivaConcept{impure function} ``count''
increments an internal counter automatically for each timestep.
Because of that, the execution of this operation cannot be simply skipped,
even if its output values are not immediately required.
B)  \DivaConcept{Global function} ``reduce'' reduces values across multiple 
MPI ranks, and its result is then used by another operation. 
In this case, this operation becomes a dependent of all 
instances of ``reduce''.
Thus, even if only one of the operations is activated 
(by a triggered action on that rank), 
all instances of ``reduce'' shall be executed
to ensure correctness.}
\label{fig:spatial-temporal}
\end{figure}


\ListCodeText{Window} is one of the \DivaConcept{impure functions} 
provided by \LANG, which creates array-like signals by 
collecting values over a range of time.
\add{Its definition (shown in~\autoref{fig:window}) takes
a target signal (\ListCodeText{field}) as the value source,
a Boolean condition (\ListCodeText{trigger}) controlling when a value is saved,
and a pair of integers (\ListCodeText{max\_size} and \ListCodeText{min\_size}) 
defining the shape of the output array.}
For example, \ListCodeText{window(true,isosurface,10,5)} 
retains up to $10$ isosurfaces at a time, 
where for \emph{every} time step a new isosurface is pushed into the window;
\add{if the number of time steps computed falls between $5$ and $10$, 
then the size of this window is the same as the time step number;
if the current time step is less than $5$, then the size of the returned array 
will be $5$ with the values in the empty space undefined.}
\add{Such an operation} 
offers support for 
time-sequence analysis, as well as 
backward feature tracking.
For example, once a feature is detected, 
the data histories associated with the feature that are within a window 
can be retained.
To provide the user with the ability to express complicated triggering conditions
on events, \LANG also adopts operations
from the field of temporal logic programming~\cite{gabbay1994temporal}. 
By abstracting commonly used Boolean operations on time sequences,
they can simplify the expression of time-varying control logic.
For example, \ListCodeText{until(x)} creates a 
Boolean signal which will be \ListCodeText{true}
until the first \ListCodeText{true} occurrence of \ListCodeText{x}.
Similarly, \ListCodeText{after(t>1700)} defines a signal whose value is 
\ListCodeText{false} until the first time where \ListCodeText{t} is above $1700$, 
and \ListCodeText{true} thereafter. Other basic operators include 
\ListCodeText{first(x),firstN(x,n)} and \ListCodeText{afterN(x,n)}.
In addition to these basic temporal logic operations, developers can also implement 
other customized \DivaConcept{impure functions}.
However they must be inherited from a dedicated base class and 
evaluated once for each timestep to ensure correctness.

\subsection{Global Functions}\label{sec:design:global}

Operations involving global synchronizations are very common for
data-distributed simulations and visualizations.
Although it is not completely obvious, it
is important \add{for lazily evaluated workflows} to have them 
explicitly handled 
to ensure program efficiency and correctness.
There are two reasons.
Firstly, in a typical classical FRP language with lazy evaluation enabled, 
computations might be accumulated until they are 
needed. If computations are parallelized, there might be
many synchronizations dispatched in a relatively short period of
time, making the system less balanced and more difficult to be optimized
toward overall throughput. 
Secondly, \add{mainstream} parallel programming interfaces typically 
\add{implement} global synchronization 
\add{as} 
matching calls (e.g., \ListCodeText{MPI\_send}, \ListCodeText{MPI\_recv}) 
\add{that should} be executed simultaneously on \add{multiple} ranks.
However, in a \add{lazily evaluated workflow}, this requisite can easily be
violated because the program control flows are data-driven; programs running
on different ranks do not always yield the same data values and control flows.


\LANG resolves this problem by classifying \DivaConcept{functions} as global or
non-global. \add{\DivaConcept{Global functions} by definition are those} 
whose execution involves global synchronizations.
\add{Thus reductions and distributed rendering} are \DivaConcept{global functions}.
\add{We refer to these as ``intrinsic'' \DivaConcept{global functions}.}
%
Additionally, if a \DivaConcept{global function} ($g$) is being 
\add{used as an input to a normally \DivaConcept{non-global function} ($f$), 
then the result ($f \circ g$) 
is also a \DivaConcept{global function}, because its evaluation would
potentially invoke $g$. 
This behavior can also be understood as building
dependencies across ranks as shown in \autoref{fig:spatial-temporal}B, 
where an activated \DivaConcept{function} ($A$) is pulling data from
an intrinsic 
\DivaConcept{global function} ($B$).}
This makes $A$ a dependent of all $B$s (e.g. on each rank).
Therefore, all the instances of $B$ should 
be evaluated, even if only one instance of $A$ needs to be evaluated.
\add{This is an example of the second type of \DivaConcept{global functions}, 
which we refer to as inherited 
\DivaConcept{global functions}.}
Different from  \DivaConcept{impure functions}, a global operation can
sometimes still be evaluated lazily, because the entire program will be
correct, as long as they are evaluated in unison on all ranks.
Thus, both \DivaConcept{pure functions} and \DivaConcept{impure functions}
can be \DivaConcept{global functions}.
\add{Though being important for performance and correctness, in \LANG, the 
computation of ``globalness'' is totally transparent to users. 
Intrinsic \DivaConcept{global functions} can be developed
using \LANG's lower level API by calling parent constructors 
(e.g., \ListCodeText{DivaImpure} or \ListCodeText{DivaPureop}) with certain flags.
If these flags are specified, parent classes will pass MPI handlers 
(initialized by the simulation) to their children.
In contrast, 
inherited \DivaConcept{global functions} can only be identified 
once the workflow has been composed.
In this case, \LANG will traverse the DAG and propagate ``globalness'' at runtime 
by marking descendants of intrinsic \DivaConcept{global functions} as global as well.
Because the definition of a signal in \LANG cannot be changed after 
compilation, its ``globalness'' only needs to be checked once for each workflow.}

%% file: contents/04-impl.tex
\section{Implementation Details}
\label{sec:impl}

\subsection{Language Parser}
\label{sec:impl:parser}



\add{The \LANG runtime translates a declarative and reactive \LANG code into 
an internal DAG representation using a parsing algorithm developed from scratch.
In particular, the parser first scans through the program and 
constructs 
a namespace object for signals.
If a signal value is defined using an inline expression, 
each operator used in the expression would create a separate namespace entry.
This design makes it easy for the system to only re-evaluate signals 
whose values have been changed. 
Once the namespace has been built, a DAG representation can be easily constructed
and validated. 
In particular, the parser checks for ``cycles''. 
A cycle means that two graph nodes are dependent on each other.
Because \LANG does not allow variable rebinding at runtime, 
this structure can produce non-executable workflows.
The parser also performs a topological sort on the DAG
and computes 
a unique evaluation order for graph nodes.
This effectively prevents the appearance of temporary inconsistencies in the 
DAG (i.e., ``glitches'').
Thus, a node can be evaluated if and only if all of its dependencies
are up-to-date.
After this, the DAG is sent to the workflow manager for execution.
\autoref{code-integration} demonstrates how the \LANG parser should be used by the
simulation in a nutshell.} 



\subsection{Workflow Manager}




The workflow manager is the component for evaluating 
the constructed DAG. 
\add{For every timestep, it is invoked once by the simulation to
execute an iteration of the workflow (see~\autoref{code-integration}).}
The workflow manager optimizes the execution following the lazy evaluation 
principle,
by deferring the evaluations of computations until their results 
are absolutely needed. 
Additionally, it also implements a value caching mechanism, 
which avoids repeatedly executing the same computation~\cite{scott2000programming}.
(Details about how different signal types are cached can be found 
in~\autoref{sec:design:impure}.)
Specifically, at each timestep, 
the workflow manager evaluates the DAG through the following four passes.

First, the workflow will try to reinitialize the execution environment
and reload dynamic libraries if necessary.
This step is intended to support 
programmers who wish to incrementally 
develop and verify their ideas,
without shutting down the simulation entirely.
\add{However, as shown in~\autoref{sec:design:impure}, if workflow 
specifications are never changed, repeated reinitialization is carefully 
avoid.}

Then, it starts evaluating all of the 
\DivaConcept{impure functions} 
that appear in the workflow. 
Even though they are internal computations, they need to be handled separately 
because they are allowed to produce side-effects, such as mutations of static states.
Processing them using lazy evaluation might lead to incorrect results.
For instance, the \ListCodeText{count} operation shown 
in~\autoref{fig:spatial-temporal}A
computes the number of timesteps that have gone by, by maintaining a counter locally;
however, if the evaluation was skipped in a previous timestep, the value 
returned from the operation would be wrong when it is needed.
Thus, to ensure the correctness of the program, these impure operations 
should be processed eagerly.

After having all impure internal components evaluated, 
\LANG will then move to impure external nodes --- \DivaConcept{actions}.
Particularly, \LANG assumes
that, in a DAG, only paths 
ending in
\DivaConcept{triggers} are meaningful, 
and all the other paths can be skipped.
For example, the process of rendering 
can be skipped for some timesteps
if the rendered image is not eventually saved in the current timestep.
However, data storage processes   
should always be executed for each timestep,
since they permanently save files to disks.
Such a prioritized evaluation is correct
because a workflow in \LANG can only produce 
side effects to the environment through \DivaConcept{actions} 
and \DivaConcept{impure functions} 
(but \DivaConcept{impure functions} are already handled in the previous step).

Finally, to avoid repeated evaluation of unused operations, 
\LANG also maintains a caching table to track pure \DivaConcept{functions}' 
most recent input parameters and values. 
If none of the inputs are changed in the current timestep, 
then the evaluation of a pure \DivaConcept{function} can be short-circuited.
Notably, because users do not directly specify the execution order for all programs,
\LANG also implements the same short-circuit mechanism internally 
for Boolean operations.
This effectively allows Boolean expressions like ``\ListCodeText{x and y}'' to 
terminate earlier without computing \ListCodeText{y} when the result of
\ListCodeText{x} turns out to be \ListCodeText{false}. 
This is a crucial feature for implementing multi-level triggers,
because these triggers are typically designed by having  
expensive analyses guarded by some 
looser but cheaper constraints
(e.g., \ListCodeText{y} is being guarded by \ListCodeText{x} in the expression
mentioned above); 
computing the Boolean value after evaluating all input variables would defeat the purpose
of multi-level triggers.

\begin{lstlisting}[float=tb,%
caption={Pseudocode demonstrating how \LANG is integrated into S3D. In 
particular, two C++ functions are introduced in S3D for initialization and \INSITU
processing. Because we do not change workflow specifications while the simulation is running,
the parsing process is written in the \ListCodeText{Init} routine. Since \DivaConcept{sources} are
essentially inputs to the workflow, they are updated (with fields passed as 
pointers) for each timestep. After that, \LANG's workflow manager is invoked to
evaluate \DivaConcept{impure functions} and \DivaConcept{triggers} respectively.
\vspace{-3em}},label={code-integration}]
map<(!string!),(!diva_source_t!)> sources; (!Language!) lang; ~...~
void Init((!int!) tstep, (!double!) time, (!double!)* ...) {
 sources["data"] = divaSourceCreate("data");
 lang.parse("case-study.diva",...);
}
void Process(int tstep, double time) {
 sources["data"]->addData("geom", ^GeomInput^(~...~));
 for (auto f : sim.fields())
   sources["data"]->addData(f.name(), ^FieldInput^(f));
 sources["data"]->commit();
 /* code executed by the workflow manager */
 lang.setSources(tstep, sources);
 for (auto v : lang.getAllImpures())  v->evaluate();
 for (auto v : lang.getAllTriggers()) v->evaluate();
}
\end{lstlisting}

%% file: contents/05-examples.tex
\section{Case Study: Anomaly Detection in S3D}\label{sec:case}
In this section, we show a real visualization and analysis workflow for combustion 
simulations implemented using \LANG. Particularly, we begin with a review of the 
simulation and the anomaly detection algorithm.
Then we explain the workflow with example codes and comparisons.
Afterwards, we describe how we establish benchmarks on the Summit supercomputer.
Finally, we conclude with benchmark results and discussion.

S3D is a scalable, reacting, compressible flow direct numerical simulation (DNS) 
solver, which is extensively used to simulate key combustion phenomena relevant to 
internal combustion and gas turbine engines~\cite{hawkes2007scalar}. The code 
solves the conservation equations for mass, momentum, energy, and chemical species 
at each grid point of a computational mesh, and over  several hundreds of thousands 
of time steps. 
At extreme scale, the volumetric data (comprised of the velocity field,
pressure, temperature, and mass fractions of about 10 to 110 chemical species) 
generated from each simulation runs into several terabytes, often overwhelming the 
I/O bandwidth and storage allowance. Therefore, the output is usually reduced by 
saving to disk at a significantly reduced temporal frequency. However, at this 
reduced frequency, the data often misses transient dynamics of exponential 
processes, such as auto-ignition that are essential for understanding the combustion 
phenomena. 
In many situations, events such as auto-ignition appear in highly localized 
regions of space and/or time. Hence, \INSITU algorithms and workflows that capture 
the events of interest are being developed to intelligently guide the saving of data 
and reduce the storage costs~\cite{aditya2019anomaly}.



\subsection{Case Overview}

%
For the demonstration and evaluation of our system, we simulate a turbulent premixed 
auto-ignition problem that is relevant to
homogeneous charge compression ignition (HCCI) and stationary gas turbine 
engines~\cite{savard2019regimes}. 
Some of the features of interest in analyzing such simulations are the conditions 
surrounding the ignition events and flame surfaces. Analyzing these conditions 
enables scientists to better understand the combustion phenomena, including the 
flame stabilization mechanisms, fuel consumption rates, and pollutant formation.  


The inception and growth of ignition kernels in 
the premixed reactants occur rapidly and, 
are often missed in coarse check-pointing of the data. 
%
\add{The anomaly detection algorithm mentioned earlier can be used to identify
an ignition kernel at its inception, as it can be defined as an extreme event.}
%
The algorithm first computes feature moment metrics (FMM) for
different sub-regions or MPI ranks of the computational domain. The FMMs are
measures in state space that contain the signatures of the extreme events. They also
quantify the contribution of different chemical components towards the ignition 
kernel formation. By comparing the FMM in the current MPI rank with the average FMM 
among all MPI ranks, we obtain a spatial anomaly metric ($m_1$);
by comparing this FMM with its values from the previous timestep, we obtain a temporal
anomaly metric ($m_2$).
If any of these two metrics are large enough (e.g., $m_1 > 0.7$ or $m_2 > 0.7$), 
an anomaly (i.e., auto-ignition) can be pronounced.

However, this anomaly detection algorithm has two major drawbacks. 
Firstly, its complexity is bounded by $O(mn^4)$, where $n$ is the number of chemical 
components and $m$ is the number of grid points in the sub-domain. 
Thus, executing the algorithm can be quite expensive.
Secondly, the algorithm can detect an auto-ignition event 
when it first occurs, but cannot be used to predict the event {\em a priori}.
Hence, 
when an auto-ignition is detected, 
features leading to it, which need to be visualized, 
would already have disappeared. 
\add{Other methods such as CEMA~\cite{lu_yoo_chen_law_2010,SHAN20123119}
or noise-tolerant trigger detection~\cite{bennett2016trigger} 
are also suitable for this problem
and are interchangeable for anomaly detection in 
our study;
but because they also suffer from the similar drawbacks, 
principles for designing
workflows with them should remain the same.}

\subsubsection{Defining Pre-Filters for Auto-Ignition}

Because the cost of running the anomaly detection is currently high,
pre-filtering (i.e., ad-hoc conditions) can be introduced 
to identify candidate regions and timesteps in advance. 
There are two well established pre-conditions of 
auto-ignition. 
First, there is a delay time associated with the formation of ignition kernels, 
which can be estimated from simple {\em a priori} calculations. 
In our particular simulation setup, the delay time is about 200 timesteps.
Second, an ignition is accompanied by ``heat release'' events, which can be computed
\INSITU. In particular, the phenomena of auto-ignition can only happen when 
there is a big enough ``heat release'', which is characterized locally by the maximum 
value within an MPI rank.
With this understanding, we formulated our pre-conditions:
\begin{lstlisting}
/* a) DIVA -----------------------------------------------*/
hr = ^max_array^(data.HeatRelease)
wait = (time > 200) && (time % 40 == 0)
adhoc, valid = hr > 1E-3, wait && adhoc
/* b) C++ Naive ------------------------------------------*/
(!double!) hr  = ^max_array^(data.data("HeatRelease"));
(!bool!) wait  = (time_i > 200) && (time_i % 40 == 0);
(!bool!) adhoc = hr < -1E-3;
(!bool!) valid = wait && adhoc;
/* c) C++ with Lazy Evaluation ---------------------------*/
(!bool!) flag = true; // define a callback function and a flag 
auto hr = [&]() { // to ensure the computation can only be
 if (flag) {      // done once per time step.
  flag = false; return ^max_array^(data.data("HeatRelease")); 
 } return 0.0; 
}
(!bool!) wait  = (time > 200) && (time % 40 == 0);
(!bool!) adhoc = valid = false; // avoid re-evaluating 
if (wait) { adhoc = hr() < -1E-3; if (adhoc) valid = true }
\end{lstlisting}
In this example, three code snapshots are displayed. Part a) is written using \LANG,
part b) is a similar implementation in C++, and c) is an optimized version in C++
following the lazy evaluation principle. Clearly, program b) is very simple, but less
efficient compared to the other programs, because in program b), the maximum ``heat 
release'' 
value will be calculated for every timestep. However, this value will be useful
only when the Boolean variable ``wait'' becomes true. In part c), lazy evaluation is
correctly implemented, at the expense of creating a callback function and using 
nested control flow.

Although we have only demonstrated one particular case here, the same 
principles apply to \INSITU analyses with multi-level filtering
mechanisms in general. These types of workflows can be programmed in \LANG more 
concisely without compromising performance due to redundant computation.


\begin{lstlisting}[float=t!,%
caption={Code comparison between \LANG and C++. Both codes compute the 
anomaly metrics by comparing the local FMM with the average FMM among all MPI ranks
and with its values from the previous timestep. In the C++ version (B), because 
the MPI operation is guarded by a if-statement, the Boolean condition needs
to be manually synchronized; however, because the manipulation of ``fmm\_win''
is local to the MPI rank, its control flow condition should not be synchronized.
This creates complexities for users.
In the \LANG version (A), global synchronization steps for control flows are 
automatically handled, which not only makes coding easier, but also reduces the
chances for mistakes.\vspace{-3em}},
label={fig:code-glob}]
/* ---- Code A -- DIVA -----------------------------------*/
features = [data.H2, ..., data.temperature]
/* feature moment metrics */
fmm = ^anomaly_detection^(features);
fmm_hst = ^window^(/*trigger=*/valid, fmm, 2, 2) 
fmm_avg = ^reduce_avg^(fmm)
/* determine anormaly */
m1 = ^anomaly_metrics^(d1=fmm, d2=fmm_avg)
m2 = ^anomaly_metrics^(data=fmm_hst)
anomaly = valid && (m1 > 0.7 || m2 > 0.7)
Trigger(valid) { ^print^("m1=" + ^str^(m1)) }
Trigger(valid) { ^print^("m2=" + ^str^(m2)) }

/* ---- Code B -- C++ with Lazy Evaluation ---------------*/
inline (!bool!) ^globalSync^(int ret,...) {
  MPI_Allreduce(&ret,&ret,/* MPI_BOR...*/); return ret; 
} ~ ------------ main program ------------ ~
auto features       = (!vector!)<(!double!)*>{...};
auto fmm            = (!deque!)<(!vector!)<(!double!)>>();
static auto fmm_avg = (!vector!)<(!double!)>(features.size());
static auto fmm_win = (!deque!)<(!vector!)<(!double!)>>();
if (^globalSync^(valid,~...~)) {
  fmm = anomaly_detection(features);
  // compute the average fmm per-feature across ranks
}
if (valid) { 
  fmm_win.push_back(fmm); 
  if (fmm_win.size() > 2) fmm_win.pop_front(); 
}
// spatial metrics (m1) & temporal metrics (m2)
(!double!) m1 = 0, m2 = 0;
if (^globalSync^(valid,~...~)) m1=^anomaly_metrics^(fmm,fmm_avg);
if (valid) m2=^anomaly_metrics^(fmm_win[1],fmm_win[0]);
\end{lstlisting}

\subsubsection{Automatic Synchronization}


To correctly detect the anomalous events, we need to compare the feature moment 
metrics across
time, and across the decomposed domain. To achieve that in a traditional language,
static storage and standard MPI synchronizations can be used as shown in 
\autoref{fig:code-glob}B.  
Since metrics $m1$ and $m2$ are only used once in a while (when variable ``valid'' 
is true), we can further optimize the code by guarding  
\ListCodeText{anomaly\_detection}, \ListCodeText{MPI\_Allreduce} and the 
manipulation of ``fmm\_win''with an if-statement.
However, part of this optimization is in fact wrong, 
as the value of the Boolean condition ``valid'' can be different on 
different MPI ranks. 
Clearly, if one of the program instance enters a different branch, 
the execution of the simulation will be blocked infinitely.
To fix this issue, we have to synchronize the Boolean condition before entering
the associated control flows (by calling the \ListCodeText{globalSync} function 
in~\autoref{fig:code-glob}B).
As we can see in practice, correctly deciding which control flow conditions to 
synchronize can 
be tedious and error-prone.
In \LANG, with the help of its built in dependency tracking, synchronizations
are automatically handled, which results in a much simpler code 
(as shown in~\autoref{fig:code-glob}A).
These details are discussed 
in~\autoref{sec:design:global}.

\subsubsection{Establishing Short-Term Memory}

The second drawback of the anomaly detection algorithm, is that it 
cannot predict anomalies
in advance. Because this algorithm is also expensive, it is 
undesirable to execute the algorithm at every timestep. 
Therefore, it is very likely that when an auto-ignition event is found, the regions 
of interests 
for visualization have already been missed.
One approach to solve the issue is to aggressively memorize all features of interests
from candidate regions in a limited RAM space.
Within these candidate regions, the anomaly detection algorithm can then be executed.
If an auto-ignition is found in a region, recorded data in the short-term,
within this region, can then be transferred to long term storage, or be used to 
trigger downstream visualization and analyses for causality studies.
This approach is practical, because data stored in short term memory are 
limited, and local to a small number of regions.
We also consider this solution superior to traditional fixed policy workflows,
because it can capture pre-ignition events correctly with much less overhead.
In \LANG we can implement this method using the \ListCodeText{window} 
function:
\begin{lstlisting}
stats_ftr_avg = ^avg_list^(features) // pre-ignition 
stats_ftr_min = ^min_list^(features) // statistics
stats_ftr_max = ^max_list^(features)
len = 40 /* record 40 steps */
recorded_avg = ^window^(stats_ftr_avg, len)
recorded_min = ^window^(stats_ftr_min, len)
recorded_max = ^window^(stats_ftr_max, len)
Trigger(anomaly) { ^save_statistics^(data=recorded_avg,~...~) 
                   ^save_statistics^(data=recorded_min,~...~) 
                   ^save_statistics^(data=recorded_max,~...~) }
\end{lstlisting}

\subsubsection{Temporal Logic to Simplify Control Flow}

\begin{figure}[tb]
\begin{center}
\includegraphics[width=1.0\linewidth]{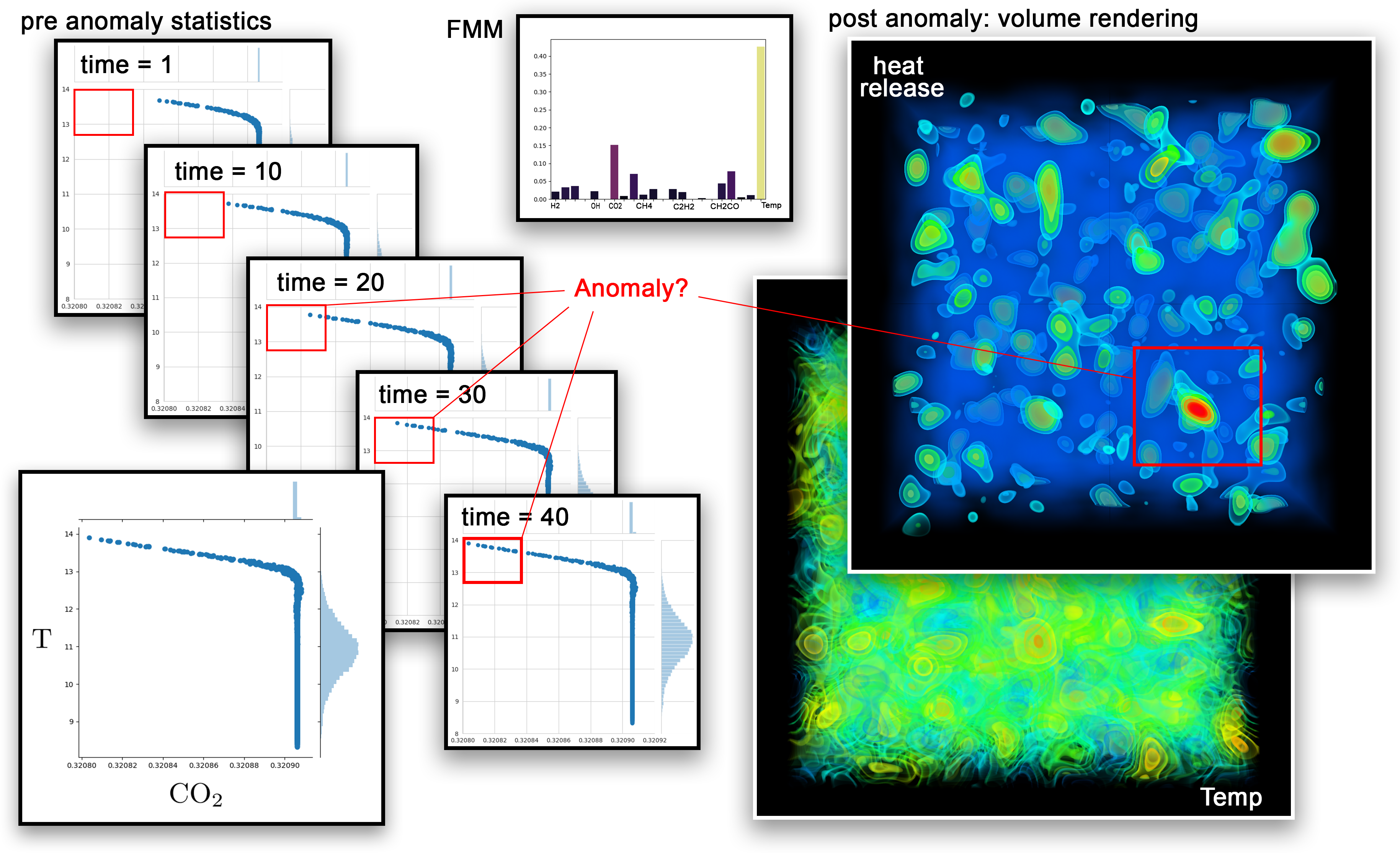}
\end{center}
\vspace{-1em}
\caption{Visualizations generated by our case study workflow.
The maximum heat release value and feature moment metrics
are used to jointly detect the phenomena of auto-ignition. To study the cause of 
auto-ignition, we visualize the statistics of raw variables computed by the 
simulation (e.g., chemical mass fractions, temperature, pressure, etc.) using joint
PDF plots (at time steps leading up to, and till the moment of ignition).
We use volume renderings of important characteristic variables (e.g., 
the heat release and temperature), to visualize the 
geometry and scales of the phenomena. We also generate histograms of the feature
moment metrics as guidance for statistical analysis. 
\vspace{-2.5em}}
\label{fig:case}
\end{figure}

One important objective for this workflow is to correctly visualize spatial–temporal
regions near the ignition kernel.
In particular, this means we should not only identify the period of time before 
the auto-ignition, but also start downstream visualizations after the ignition (as 
illustrated by~\autoref{fig:case}).
Formally, the pre-ignition period is defined as the time interval between
the appearance of the first pre-filtering condition till the appearance of anomaly; 
while the post-ignition interval starts with the anomaly and lasts for a fixed
period of time.
However, because the pre-filtering condition is data-driven, 
it will naturally fluctuate as the data changes, 
making it unsuitable for defining continuous time intervals.
In \LANG, these sort of problems are handled by built-in
temporal logic functions (as discussed in \autoref{sec:design:impure}).
Particularly, \LANG provides functions like ``switch''\footnote{The ``switch'' 
function is analogous to the behavior of a light switch, which can be turned on by 
the first ``on'' condition if it is currently off, and turned off by the first 
``off'' condition if it's currently on.} and ``countN'' to convert discrete events 
into intervals. Therefore, we can easily define our pre-ignition and post-ignition 
events as in the following example:
\begin{lstlisting}
pre_anomaly  = ^switch^(on = valid, off = anomaly)
post_anomaly = ^countN^(since = anomaly, n = 10)
/* render the heat release after anomaly for 10 steps */
vol = ^volume^(field = data.HeatRelease,~...~)
Trigger(post_anomaly) { ^save_ppm^(vol, "img-" + ^str^(time)) }
\end{lstlisting}

\begin{figure}[tb]
\begin{center}
\includegraphics[width=\linewidth]{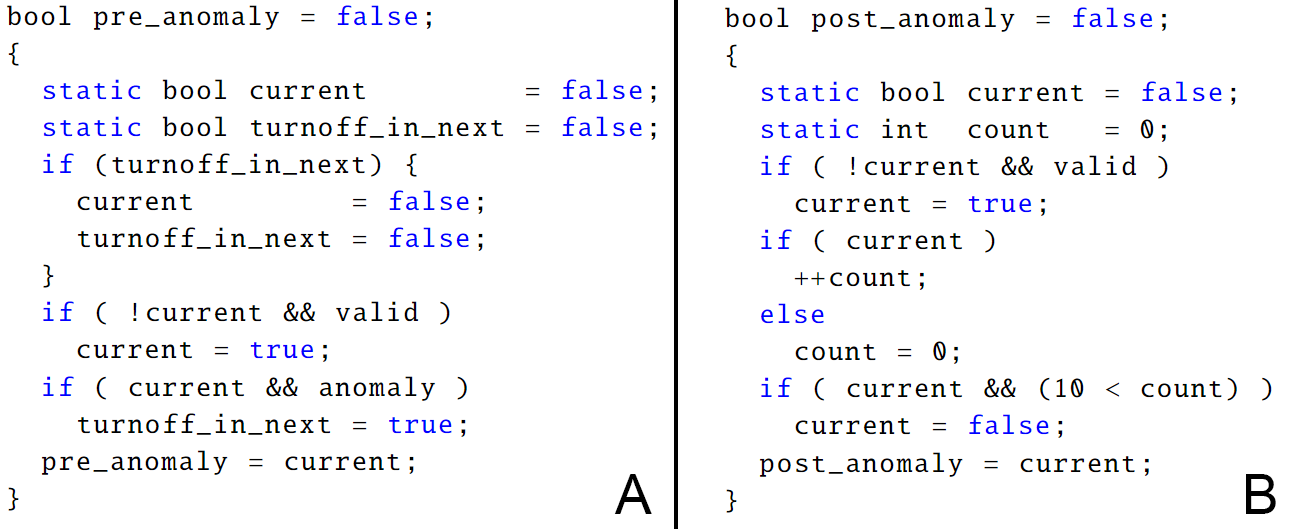}
\end{center}
\vspace{-1em}
\caption{C++ implementations of the pre-anomaly condition and the post-anomaly 
condition. The pre-anomaly phase is defined as a continuous time interval 
from the moment that variable ``valid'' becomes true, to the moment that variable 
``anomaly'' becomes true. The post-anomaly phase is defined as a fixed length 
interval since ``anomaly '' becomes true. As we can see, defining continuous time 
intervals from time-varying values can be tedious in traditional languages such as 
C++.}
\label{fig:code-temp}
\end{figure}


Although these temporal functions look very simple in \LANG, they can be fairly hard
to implement in traditional languages like C++ or Python (examples are shown 
in~\autoref{fig:code-temp}). 
This is not only because the manipulation of static storage can be complicated, but 
also because those functions should only be executed once globally for each timestep.
In other words, these functions can only be implemented 
with separately maintained local storage (as \LANG does internally).

\begin{table}[t]
\caption{Benchmark Results.}
\label{tab:benchmark}
\scriptsize\centering
\resizebox{\columnwidth}{!}{
\begin{tabu}{*{1}{l}*{6}{r}}
\toprule
Grid Size
& $128^3$
& $128^3$
& $256^3$
& $256^3$
& $512^3$
& $512^3$
\\
Size per Rank
& $32^3$ 
& $16^3$ 
& $32^3$ 
& $16^3$ 
& $32^3$
& $16^3$\\
\midrule 
Nodes
& 2 
& 16
& 16
& 128
& 128
& 1024\\
Proc per Node & 32 & 32  & 32 & 32 & 32 & 32 \\
GPU per Node & 1 & 1  & 1  & 1 & 1 & 1 \\
\midrule 
$\text{Time}_{\text{Ref}}$   (s) & 1677 & 769 & 2321 & 980 & 2959 & 1444 \\
$\text{Time}_{\text{\LANG}}$ (s) & 1639 & 779 & 2194 & 959 & 2607 & 1388 \\
\% Difference & -2.3\%  & 1.3\%  & -5.5\%  & -2.1\% & -12.0\% & -3.9\%     \\
\bottomrule
\end{tabu}
}
\vspace{-2em}
\end{table}

\subsection{Benchmark}

To effectively estimate the performance of our system, we compare our adaptive 
workflow implemented using \LANG with a reference implementation of the 
workflow written directly using C++.
We optimized this reference implementation following almost the same principles
we used to optimize \LANG, except we did not implement lazy evaluation for trivial computations,
such as simple arithmetic.
We believe our reference implementation is an efficient implementation of the 
workflow when no additional parallelism layers are involved.
For modular operations, such as the computation of the feature moment
metrics, identical implementations are used in both workflows.

To verify our assertions, we compiled both implementations \add{with a CPU-based
S3D} using identical compiler settings 
(PGI compiler in the default ``Release'' mode configured by CMake),
and benchmarked them across 6 different configurations on the Summit supercomputer.
For each configuration, we ran both implementations once.
For each configuration, we placed 32 MPI ranks on each compute node
with 2 IBM POWER9 22-core CPUs,
and requested 1 NVIDIA Volta V100 GPU for each compute node.
The GPUs were used by the GPU-based volume rendering library
integrated in \LANG.
The number of compute nodes we requested for each configuration 
can be found in~\autoref{tab:benchmark}.
%
\add{To qualitatively assess our implementation,  
we measured the overall workflow processing time for each run
by summing the workflow time of each timestep.
In particular, we profiled each run for exactly 220 steps, starting 
from a middle timestep checkpoint, and we guaranteed that: first, runs of the same
configuration shared the same checkpoint file; second, auto-ignition was happening 
by the end of each run; third, visualizations and statistics produced by runs
from the same configurations agreed with each other.}

Our results are summarized in \autoref{tab:benchmark} with all timings measured in 
seconds.
\add{We found that, for most of} the configurations,
\add{the two implementations indeed have similar performance, with
low percentage differences\footnote{
~Percentage Difference~(\%) =
$\frac{\text{Time}_{\text{\LANG}} -
\text{Time}_{\text{Ref}}}
{\text{Time}_{\text{Ref}}} \times 100$}
($<\pm6\%$).
The $512^3$-$32^3$ configuration was the only exception. 
For this configuration, 
we found that \LANG was able to compute FMM
faster consistently (by about 10 seconds). 
However, we did not observe the same 
phenomenon with other configurations, which suggests that
the discrepancy might not be due to \LANG.}

%% file: contents/08-discussion.tex
\section{Discussion}

There are other potential uses of \LANG beyond programming dynamic \INSITU workflows.
For example, the current implementation of \LANG does not directly support 
\textit{in transit} workflows. But we could implement a workaround by
developing customized
\DivaConcept{sources} and \DivaConcept{actions} 
using \textit{in transit} libraries such as ADIOS~\cite{ADIOS}
\add{and having two workflows running alongside each-other either synchronously or asynchronously.}
\add{In particular, this would require one workflow to be running \INSITU 
and writing data using an ADIOS \DivaConcept{action}, 
and the other workflow to be running
separately and receiving data using a connected ADIOS \DivaConcept{source}.}
With a focus on expressiveness and code portability,
\LANG can also be used as a thin layer to enable 
declarative and reactive programming
on existing frameworks like ALPINE~\cite{Ascent} and SENSEI~\cite{SENSEI}.
This \add{would also} allow users to port codes across different 
platforms and frameworks.
\add{For example, one could use the native \LANG implementation 
and a GPU workstation to develop and debug a workflow, 
and then directly deploy the program to a supercomputer that features
a CPU-based rendering infrastructure such as 
VisIt-OSPRay~\cite{Wu_VisItOSPRay_2018}.}
%

\add{There are several limitations for \LANG. }
\add{First, \LANG currently does not support programming loops and functions
directly in its declarative interface. If these are needed,
the low-level API needs to be used.}
%
%
\add{Second, \LANG's current implementation 
has almost no restrictions for module 
implementations. Bugs in the extension (e.g., accessing MPI
handlers from \DivaConcept{non-global functions}) can be very hard to find 
and can lead to unpredictable results as workflows are highly dynamic.}
\add{Third, \LANG currently allows \DivaConcept{global functions} to 
communicate directly through MPI.}
\add{While modules can also exploit heterogeneous parallelism internally 
(e.g., through VTKm worklets), there are remaining challenges towards optimizing
data and resource management (e.g. through dynamic resource allocation).
To solve this problem, a more generic low-level data parallel programming 
environment such as Legion~\cite{Legion} would be needed.}
\add{Fourth, integrating \LANG into simulations currently}
requires changes to be made to the 
simulation code
\add{because the simulation would be in charge of creating \LANG
\DivaConcept{sources}.}
This could result in many different customized versions being maintained for only 
slightly different purposes. 
\add{Thus a simpler and more generic way for the simulation side integration 
would be very helpful.}
\add{Finally, \LANG's current design prohibits variable re-definition. 
Thus, data dependencies in a \LANG workflow can not be modified once
compiled. This assumption greatly reduces the complexity of \LANG in design; 
however, it also prohibits the use of triggers on internal \DivaConcept{functions}.
Finding a way to achieve this could be an interesting direction for future work.}

 
\section{Conclusion}
As we enter the age of extreme scale supercomputing and Big Data, the support of 
\INSITU data analysis and visualization becomes indispensable to application 
developers and domain scientists. 
The introduction of \LANG, 
a declarative and reactive programming environment,
overall makes adaptive \INSITU workflow development a simpler process. 
We find  the key benefits that it provides include: 
more autonomy between developers, 
modularity of workflow components, 
extensibility through the back-end runtime system, 
protection against logical programming errors by using implicit control flow 
execution, and
a more results-oriented paradigm that better reflects end goals.
As \LANG matures, it shall continue to be refined and extended to support a wide 
range of applications.